\documentclass{ws-procs9x6}

\begin{document}

\title{The cosmological evolution of the environments of powerful radio galaxies}

\author{J.A.Goodlet \& C.R.Kaiser}

\address{School of Physics and Astronomy\\
Southampton University\\
SO17 1BJ\\
United Kingdom\\ 
E-mail: jag@astro.soton.ac.uk}


\maketitle

\abstracts{We present the results from the 
analysis of 26 extragalactic radio sources of type FRII which were
observed with the VLA at 5 GHz and around the 1.4 GHz band. The
sources were selected to have redshifts in the range $ 0.3<z<1.3$,
radio powers between $6.9 \times 10^{26} {\rm WHz^{-1}}<P_{151 {\rm
MHz}}<1.3 \times 10^{28} {\rm WHz^{-1}}$ and angular size $\theta \ge
10''$. We found that the depolarisation and the rms variations in the
rotation measure increased with redshift. The flux values obtained from
the observations were used to derive by means of analytical modelling
the jet--power, density of the central environment, age of the source
and its lobe pressure and the results were then compared with the
observations. We find no significant correlations with the density
parameter suggesting that the depolarisation and the rms variations in
the rotation measure are indicative of the environment becoming more
disordered rather than denser.  The age and size of a source are
correlated and both were found to be independent of redshift and
radio--power. Jet--power strongly correlated with the
radio--power. The lobe pressure was found to be anti--correlated with
size which could explain why there are no sources beyond a few Mpc in
size. We found no significant correlation between size and density
which demonstrates that the sample is a fair representation of the
population.}

\section{Observations}
\label{obser}
To break the degeneracy effect found between radio--power and redshift
in previous studies we defined 3 subsamples of sources chosen from
the 3C and 6C/7C catalogues.  For full details on the sample selection
and the data reduction see Goodlet {\it et~al.}\cite{me}, hereafter
G04. Spectral index, rotation measure and depolarisation were
derived from the observations and were averaged over individual lobes.
We define the spectral index, $\alpha$, by $S_{\nu} \propto
\nu^{\alpha}$. The difference in $\alpha$, DM and RM between 
the two lobes of an individual source is given by d$\alpha$, dDM and
dRM respectively. The rms variation of the rotation measure is defined
by $\sigma_{RM}$.


\section{Theoretical modelling}
The large scale structure of FRII sources is formed from twin jets
emerging from a central AGN buried inside the nucleus of the host
galaxy. The jets propagate in opposite directions from the core of the
source and end in strong shocks. The jet material inflates a lobe
surrounding the jet, which drives a bow shock into the surrounding
medium. Kaiser \& Alexander\cite{ka97} (KA) showed that in a purely
dynamical model of FRII evolution, the bow shock and lobe grow
self--similarly. The KA model assumes a power--law for the external
density distribution, $\rho = \rho _0 \left( r / a_0 \right)
^{-\beta}$ and a constant rate of energy injection into the lobe,
Q$_\circ$. The length of the jet, D$_{lobe}$, grows with time
proportional to $t^{3/\left( 5 - \beta
\right)}$.  Kaiser {\it et~al.}\cite{kat97} (KDA) added synchrotron radio emission to the
dynamical model of KA. The model self-consistently incorporates the
effects of all relevant energy losses on the relativistic electrons
giving rise to the synchrotron emission. This allows us to calculate
the total radio luminosity $P_{\nu}$ of the lobe.  The density of the
environment is parameterised by $a_\circ^\beta\rho_\circ$ and is
related to the D$_{lobe}$, Q$_\circ$ and the age of the source. Using
the flux measurements at our 3 observing frequencies and the model of
KDA, we constrained the lobe pressure and external density of all
sources in our samples. Q$_\circ$ and the source age were then
calculated from the best fit parameters.

\section{Results}
DM and $\sigma_{RM}$, which are indirect measurements of density,
increase with redshift. Unfortunately this was not observed in the
density parameter $a_\circ^\beta\rho_\circ$. DM and $\sigma_{RM}$ may
increase with redshift because the environments are more disturbed
rather than denser. This may fit with observations indicating a higher
degree of distortions of the large scale radio structure of objects at
high redshift objects. Spectral index was found to be independent of
all the other source properties. Larger sources are older, as expected
for ram pressure confined jets. Sources with larger jet--powers are
more luminous. Lobe pressure correlates strongly with lobe size and
the radio luminosity of the source.
%
%
%
%

\end{document}